# Towards the global vision of engagement of Generation Z at the workplace: Mathematical modeling

**Radoslaw A. KYCIA**[1]

Department of Mathematics and Statistics

Masaryk University, The Czech Republic

Faculty of Materials Science and Physics

Cracow University of Technology, Poland

email: kycia.radoslaw@gmail.com

**Agnieszka NIEMCZYNOWICZ**

Faculty of Mathematics and Computer Science

University of Warmia and Mazury in Olsztyn, Poland

email: niemaga@matman.uwm.edu.pl

**Joanna NIEZURAWSKA-ZAJAC**

Faculty of Finance and Management

WSB University in Toruń, Poland

email:  j.niezurawska@interia.pl

**Abstract:**  Correlation and cluster analyses (k-Means, Gaussian Mixture Models) were performed on Generation Z engagement surveys at the workplace. The clustering indicates relations between various factors that describe the engagement of employees. The most noticeable factors are a clear statement about the responsibilities at work, and challenging work. These factors are essential in practice. The results of this paper can be used in preparing better motivational systems aimed at Generation Z employees.

**Keywords:** Employee Engagement,  Generation Z, Mathematical Modelling,  Principal Component Analysis, Gaussian Mixture Model, k-Means Clustering

# 1. Introduction

The engagement of employees at the workplace is one of the main ingredients for company growth. Therefore, the motivational systems that encourage engagement in the staff can significantly boost the realization of development aids.

With the births ranging from the late 1990s till 2010s, the persons from Generation Z started or soon will start their first jobs in companies. High productivity of employees from this generation can be achieved by crafting a proper motivation system. Such a system must also be designed to tie the employee with the company since otherwise, the experience will be lost during the work rotation.

The importance of employee engagement to organizational outcomes is paramount. The emphasis on work engagement is progressively predominant among experts and academics, since it captures a significant part in work behavior, that is, how much energy, attention, and focus they put into work, e.g., Kahn (1990). Work engagement has been a concept that is not well understood in motivational processes, see Denunzio and Naidoo (2018). As a result, leadership has a significant impact on employee engagement within a company. Employee engagement is also related to the relationship between the company's leaders and followers, see Seijts, Woodwark and Savage (2018). The majority of previous research has focused on individual or group engagement, with little effort made to comprehend the employee engagement process across organizational levels. In this article we focus on individual engagement.

In the article we focus on Generation Z employee engagement. We will identify the main ingredients that can improve engagement and help to enhance current motivation plans targeted at Generation Z. The paper is organized as follows: In the next section the theoretical background is presented, and the following section provides the description of the data and the methods of analysis. Then the results are presented using three different ways: the correlation analysis, k-Means clustering algorithm, and the Gaussian Mixture Model. In the Appendix the details of the survey are given.

# 2. Theoretical background

Work engagement is described by Kahn (1990) as an employee's psychological appearance in his or her work position. Work engagement is also defined by Rothbard (2001) as the degree of work absorption. Employee engagement is also referred to as 'work engagement', which refers to how enthusiastic an individual employee is about his or her employment, see Farndale and Rich (2018).

Personal engagement and personal disengagement were discussed by Kahn (1990). He described personal engagement as the use of members of the company in their employee roles and personal disengagement as a physical, cognitive, or emotional withdrawal from employee roles, see Kahn (1990). Kahn was the first to argue that people who are emotionally engaged in their work invest positive emotional and cognitive energy in achieving their goals, e.g., Armstrong (2006).. According to Rich, Lepine and Crawford (2010), engagement happens when members of an organization put their potential and resources into physical, cognitive, and emotional action. Employee engagement is described by the American Society of Human Resource Management as an employee's relationship with their job, commitment to the company, and loyalty to their employer, see Rich, Lepine and Crawford (2010). Given its clear connection to significant attitudinal and behavioral results, a deeper understanding of the organizational engagement process is advantageous, see Christian, Garza, and Slaughter (2011); May, Gilson and Harter (2004); Rich, Lepine and Crawford (2010). When considering the effect of policies on work engagement, Smith and Dumas (2007), and reflecting on multiple policies, the essence of the temporal change will be significant, see Perry-Smith and Blum (2000). Besides, there is evidence in the literature that this organizational-level engagement mediates the relationship between motivational activities and organizational performance, e.g., Thurgood, Smith and Barrick (2013).

# 3. Data and preliminary analysis

## 3.1 Survey and statistics

Empirical data were collected via an online survey entitled 'Motivating and rewarding employees in relation to their engagement and loyalty at work as seen by Generations X, Y, Z', conducted in Poland in 2018. The survey was comprised of seventh parts. For our study we focus on the analysis of the first and fifth parts. The first part was concerned with respondents' socio-demographics, namely gender, age, education status, size of your company/institution, and seniority. The second one, according to Armstrong's researches in Armstrong (2006), consisted of twenty items (Q5.1 – Q5.20) related to engagement at your workplace. All the items are measured on a 5-point scale ranging from 1 (unimportant) to 5 (very important). The questionnaire for this part is presented in the Appendix. The proprietary survey was sent by various Internet Platforms using the quatrix tool QMETRICS. The sample consists of 200 participants solely from Generation Z. The structure of the responders is presented in Figs. 1-4. The sample was overpopulated by females and employees with higher education. As for respondents' seniority structure, the beginning and junior employees are highly represented over senior staff.

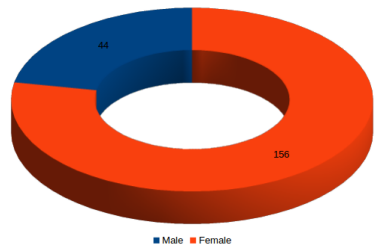

Fig. 1. The gender structure of respondents.

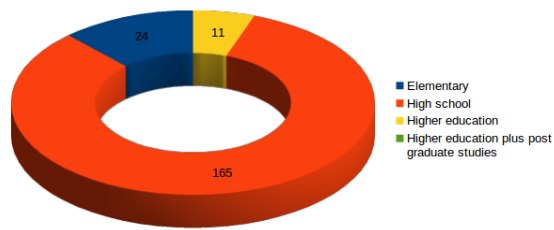

Fig. 2. Education structure of respondents.

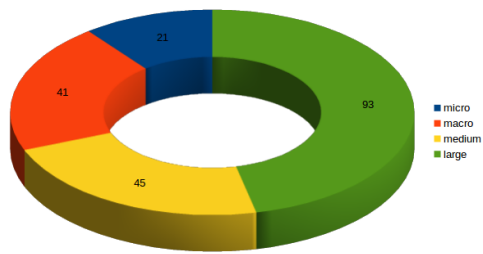

Fig. 3. Size of the company.

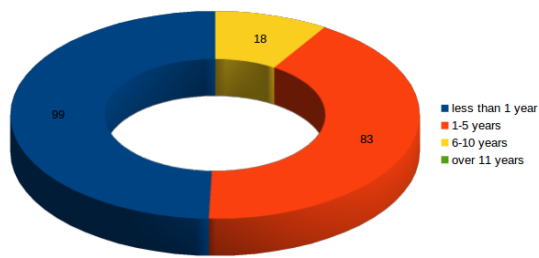

Fig. 4. Seniority.

## 3.2 Methods of data analysis

### 3.2.1 Reliability test

The data were first analyzed against internal consistency, with the standard Cronbach's alpha test. A common approach is that the acceptable values of Cronbach's alpha should fit in the range from 0.7 to 0.95 to consider the data to be reliable, as described in Cho (2016). In the case of Q5.1-Q5.20 the alpha value is 0.94, which proves its internal consistency.

### 3.2.1 Outline of the methods

The analysis aims at discovering the structure of the data and relations between the questions Q5.1-Q5.20. The state of the art methods of cluster analysis were used:

1) **Principal Component Analysis** (PCA) results in constructing a linear combination of data so that eigenvectors of covariance matrix describe the new variables. Usually, only some of the coordinates that their cumulative explained variance ratio is above a threshold are selected. Therefore PCA is used to decrease the dimensionality of the data without sacrificing too much variance/information of the original data.

2) **k-Means cluster algorithm** that aims to associate data with k clusters with centers given by the means of the points within the cluster.

3) **Gaussian Mixture Model (GMM)** in which we search for splitting the data into clusters described by a weighted combination of Gaussian distributions – one Gaussian for each cluster. The effectiveness of the splitting can be measured by minimization of the Bayesian Information Criterion.

## 3.3 Analysis pipeline

The data are treated as classless, and a typical cluster analysis is performed. The data are analyzed using Scikit-Learn Python Library, version 0.20.0, see Scikit-Learn Library Documentation, installed on Ubuntu 16.04 LTS.

The processing pipeline consists of several steps (cf. Fig. 5):

1) (Spearman) Correlation analysis of questions. This part gives all possible relations between questions, however, it should be interpreted with care since the high correlation between questions not always means the linear relation.

2) Principal Component Analysis (PCA) on standardized data and selecting an optimal number of Principal Components that explains more than 70% of the total variance.

3) Selecting the optimal number of clusters using the k-Means classifier on the standardized PCs selected in the previous step. The optimal number of clusters is inferred from the elbow method.

4) For the fixed number of clusters in the previous step, use the Bayesian Information Criterion (BIC) to select optimal Gaussian model in Gaussian Mixture Model (GMM) from available in Scikit-Learn Library Documentation: 'spherical', 'tied', 'diagonal', and 'full'.

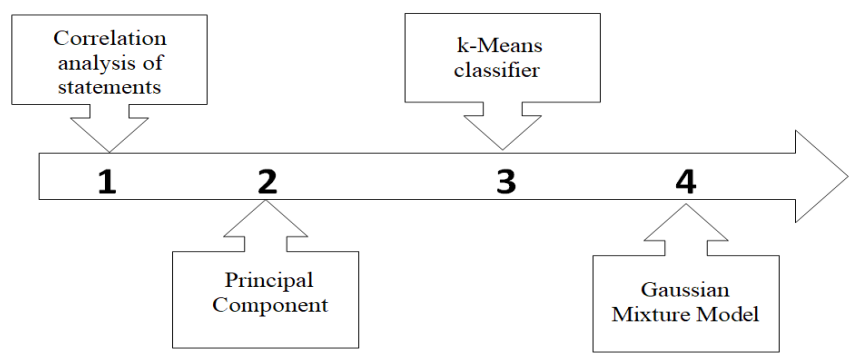

Fig. 5 Data processing pipeline.

In the following section the results of the analysis are presented.

## 4. Results

The detailed analysis along the steps described in the previous section will be presented below.

**4.1 Correlation analysis**

The Spearman correlation matrix is presented in Fig. 5.

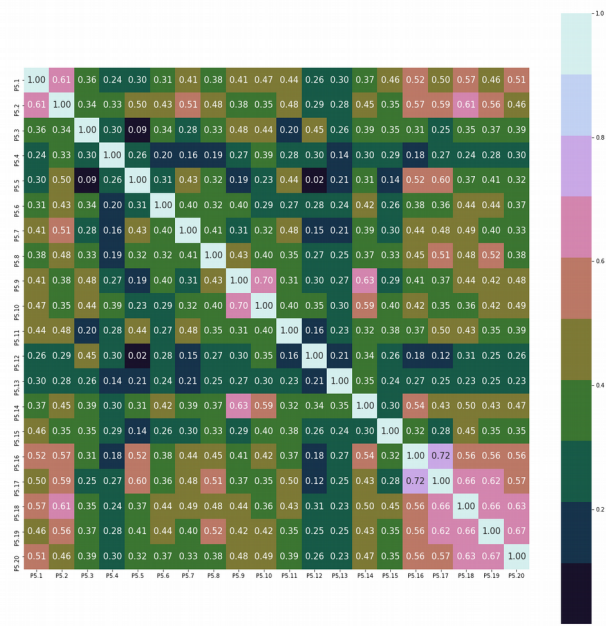

Fig. 5. Spearman correlation coefficients between questions.

The positive correlations above 0.5 suggest the existence of the following groups in the data, which are also suggestions for designing motivation systems.

- The first group - clusters expression of general satisfaction from work:
    - Q5.1 (I'm very satisfied with the work I do)
    - Q5.2 (My job is interesting)
    - Q5.16 (I think this organization is a great place to work)
    - Q5.17 (I believe I have a great future in this organization)
    - Q5.18 (I intend to go on working for this organization)
    - Q5.19 (I am happy about the values of this organization – how it conducts its business)
    - Q5.20 (The products/services provided by this organization are excellent)
- The second group - connects learning opportunities and the level of interests from work:
    - Q5.2 (My job is interesting)
    - Q5.7 (I get plenty of opportunities to learn in this job)

- The third group - connects challenging work and satisfaction with excellent prospects for the future:
    - Q5.5 (My job is challenging (sets new goals, is prospective))
    - Q5.16 (I think this organization is a great place to work)
    - Q5.17 (I believe I have a great future in this organization)
- The fourth group - connects the quality of tools and facilities with the values and with prospects for the future:
    - Q5.8 (The facilities/equipment/tools provided are excellent)
    - Q5.17 (I believe I have a great future in this organization)
    - Q5.19 (I am happy about the values of this organization – how it conducts its business)
- The fifth group - clusters the relations with the boss:
    - Q5.9 (I have a lot of support from my boss)
    - Q5.10 (My boss recognizes my work)
    - Q5.14 (I like working with my boss)
- The sixth group - indicates that the relationship with the boss is the main ingredient of satisfaction from work:
    - Q5.14 (I like working with my boss)
    - Q5.16 (I think this organization is a great place to work)

**4.2 Principal Components Analysis**

The dimensionality reduction of the data was attained by PCA analysis of standardized data. The explained variance ratio plot is presented in Fig. 6.

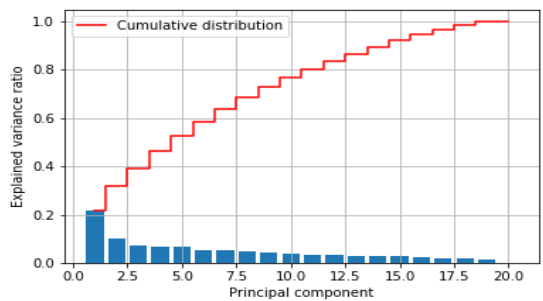

Fig. 6. PCA analysis.

The analysis shows that the explained variance ratio at the level of 73% is achieved for 9 components, which will be assumed in the following subsections.

**4.3. Optimal number of clusters by k-Means algorithm**

Standardized 9 Principal Components of the previous step were selected to cluster analysis using the k-Means algorithm. The standard elbow-type argument (see Fig. 7) suggests that the optimal number of clusters is 8.

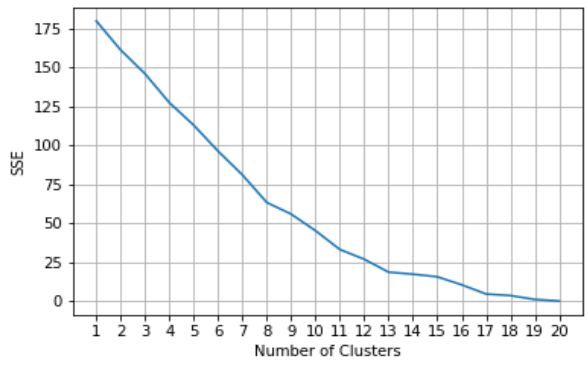

Fig. 7. The Error Sum of Squares vs the number of clusters for the k-Means algorithm.

The clusters for k=8 are presented in Tab. 1, and its projection to the space spanned by the first three PCs is presented in Fig. 8.

*Table 1: Clusters for the k-Means algorithm with 8 clusters.*

| Cluster number | Questions in clusters |
|---|---|
| 0 | Q5.2, Q5.5, Q5.11 |
| 1 | Q5.9, Q5.10, Q5.14 |
| 2 | Q5.1, Q5.16, Q5.20 |
| 3 | Q5.3, Q5.12, Q5.13, Q5.15 |
| 4 | Q5.17, Q5.18. Q5.19 |
| 5 | Q5.6, Q5.8 |
| 6 | Q5.7 |
| 7 | Q5.4 |

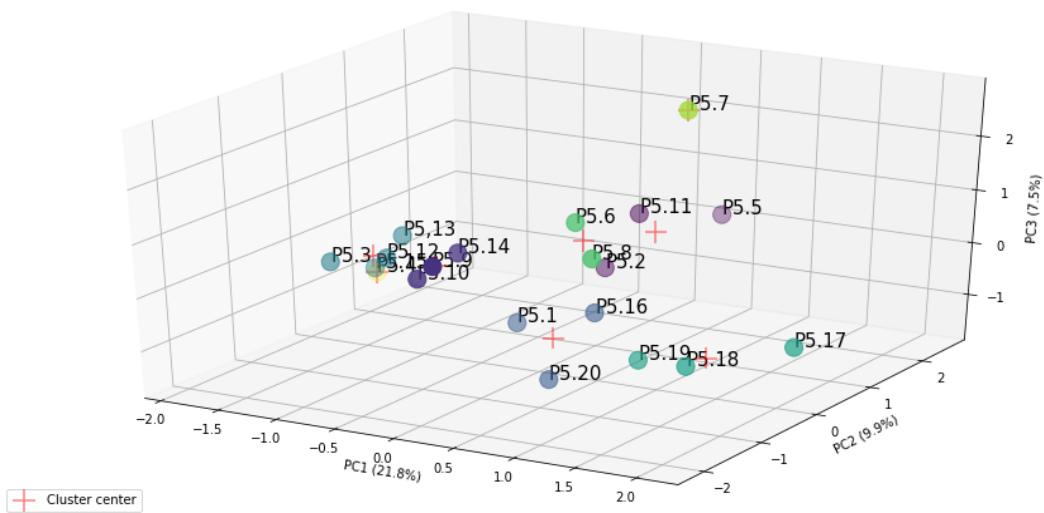

Fig. 8. Cluster analysis for k-Means algorithm and 8 clusters. The results are projected onto the subspace spanned by the first three PC.

The aggregation suggests the following connections:

- Class 0 – the interest in the job is connected with challenging tasks at work and gaining experience for the future. This connection is not reflected in the correlation analysis.

- Class 1 – represents the relationship with the boss. The questions in the class are also highly correlated.

- Class 2 – connects the satisfaction from the work and workplace with the quality of company products. The questions in the class are also highly correlated.

- Class 3 – aggregates questions that show that the clear statement of work responsibilities helps employees manage at work, have good relationships with colleagues, and attain a balance between work and personal life. This connection is not reflected in the correlation analysis.

- Class 4 – connects the intent to work for the organization with the way the company conducts the business. The questions in the class are also highly correlated.

- Class 5 – represents the connection between freedom at work with the quality of facilities and equipment at the workplace. This connection is not reflected in the correlation analysis above.

- Class 6 – opportunities to learn at work do not aggregate.

- Class 7 – an initiative to do work well is unrelated to other questions. This is also true in correlations analysis.

**4.4. Gaussian Mixture Model**

The final analysis is a Gaussian Mixture Model (GMM) for 8 clusters on standardized PCs. The type of the model is selected using the Bayesian Information Criterion. The results of the analysis are presented in Fig. 9.

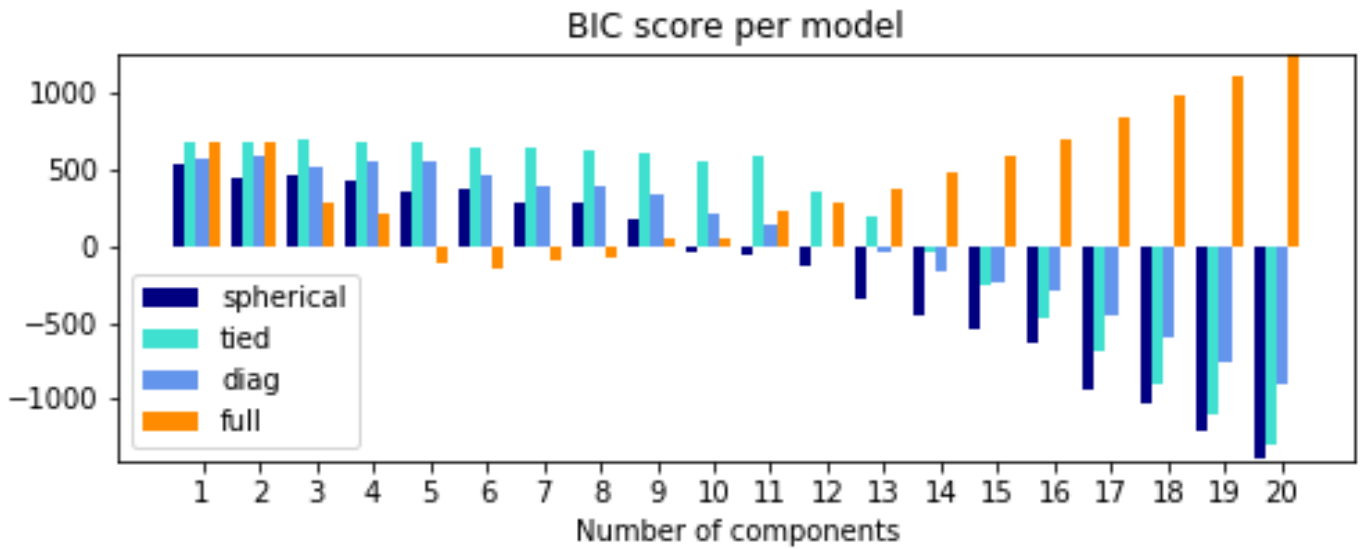

Fig. 9. BIC analysis for GMM.

For further analysis 8 *'full'* (all covariances are independent) Gaussians are used. The classes are presented in Tab. 2, and Fig. 10.

*Table 2: Clusters for GMM for 8 clusters.*

| Cluster numbers | Questions in clusters |
| --- | --- |
| 0 | Q5.13 |
| 1 | Q5.17, Q5.18, Q5.19, Q5.20 |
| 2 | Q5.1, Q5.2, Q5.5, Q5.16 |
| 3 | Q5.9, Q5.10, Q5.14 |
| 4 | Q5.7 |
| 5 | Q5.6 |
| 6 | Q5.3, Q5.4, Q5.8, Q5.12, Q5.15 |
| 7 | Q5.11 |

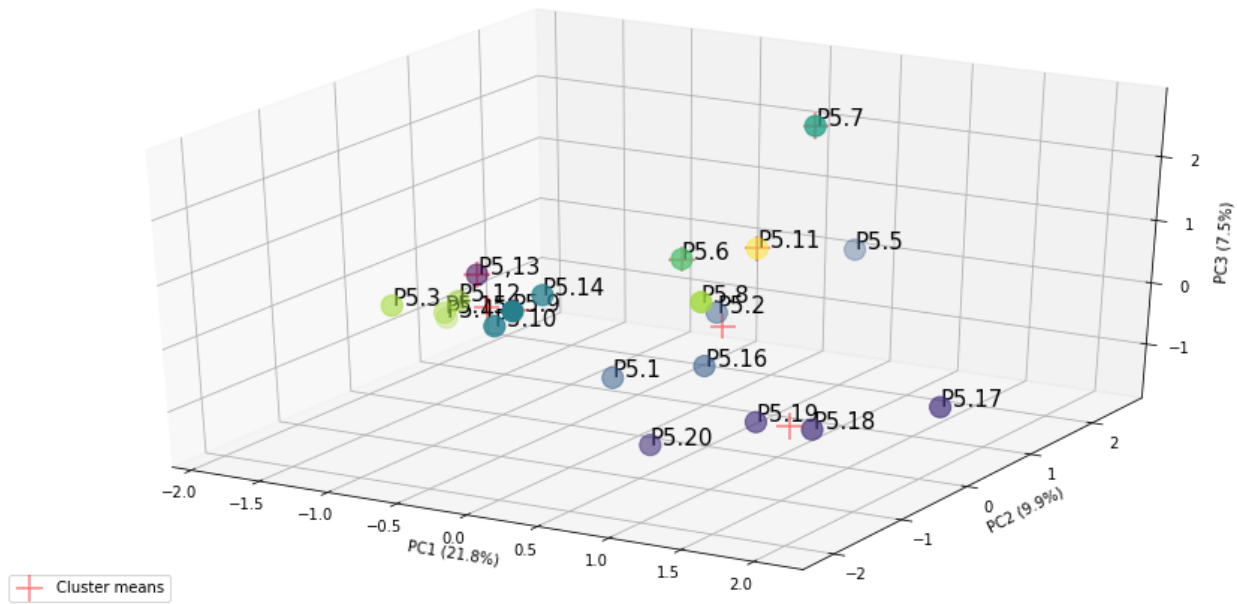

Fig. 10. GMM analysis for 8 clusters projected to the space spanned by the first three Principal Components. The centers for Gaussian distributions were marked.

We provide the discussion of the clustering:

- Class 0 – the balance between life and work is unrelated to other questions in the GMM model.

- Class 1 – relates the plans to work for the company in the future and the quality of products, and the conduction of the business. This connection is also reflected in correlations analysis and the k-Means approach.

- Class 2 – shows that satisfaction in a company is connected with interesting and challenging work. This result is also present in the previous analyses.

- Class 3 – is related to relations with the boss.

- Class 4 – learning opportunities are not connected with other questions, as it was also indicated in the previous sections.

- Class 5 – the freedom in work is unrelated to the other questions in the GMM model.

- Class 6 – the clear statements about work responsibilities are connected with the intention to show initiative at work and ease to keep up with the work and colleagues. The quality of facilities and equipment is also involved. GMM model shows that previously unrelated initiatives to do work well are related to clear responsibilities at the workplace.

- Class 7 – represents gaining experience and is unrelated in GMM to the other questions.

## 5. Conclusions

The above analysis gives an abundance of information about relations between different factors present at the workplace and employee engagement. The most significant relations that reoccur in the above analyses are:

- Interesting and challenging work is connected with employee satisfaction.

- Clear statements about the employee's responsibilities are a key to work-life balance, good relations with colleagues, and impact (in GMM model) on efficient work.

- The relation to the boss impacts the great atmosphere at work in correlation analysis; however, it is not related to other questions otherwise.

- The quality of products and how the company conducts the business are related to connecting employees' future with this company.

The other relations that are present in only one model require further detailed investigations.

These results can help to improve the motivation systems for GenZ.

## Appendix. Questionnaire

Dear Participants,

I would like to invite you to participate in research devoted to selected issues in managing generational diversity. The goal of the research is to identify effective tools for motivating of employees from the perspective of representatives of Generation Z. When you are answering the questions in the survey, please select only one answer on the given measuring scale. If you don't know the answer to a particular questions, just please leave it blank. The questionnaire is anonymous. The results will only be used for scientific research. I'd like to thank you very much for participating in the research.

**Q0**: *Background informations*
Gender

| | |
|---|---|
| Female | ☐ |
| Male | ☐ |

I. Age (born between)

| | |
|---|---|
| 1995-2000 (Generation Z) | ☐ |
| 1994-1980 (Generation Y) | ☐ |
| 1979-1965 (Generation X) | ☐ |
| 1964-1946 (Baby Boomers) | ☐ |
| 1925-1945 (Traditionalists) | ☐ |

II. Educational level

| | |
|---|---|
| elementary | ☐ |
| high school (matura exam or equivalent) | ☐ |
| higher education | ☐ |
| higher education plus post graduate studies (or doctoral studies) | ☐ |

III. Size of your company/institution

| | |
|---|---|
| micro | ☐ |
| small | ☐ |
| medium | ☐ |

large ☐

IV.         Seniority

less than 1 year ☐

1-5 years ☐

6-10 years ☐

over 11 years ☐

Directions: Please indicate your answer by tick the appropriate number. The higher level the more important the modern systems and concepts of remuneration and motivation.

(Note: 1→unimportant; 2→not so important; 3→ moderately important; 4→ important; 5→very important)

**Q5**: *Do you agree with the statements below about your engagement at your workplace?*

|  | 1 | 2 | 3 | 4 | 5 |
|---|---|---|---|---|---|
| 1. I'm very satisfied with the work i do | ☐ | ☐ | ☐ | ☐ | ☐ |
| 2. My job is interesting | ☐ | ☐ | ☐ | ☐ | ☐ |
| 3. I know exactly what I'm expected to do | ☐ | ☐ | ☐ | ☐ | ☐ |
| 4. I am prepared to show initiative to do my work well | ☐ | ☐ | ☐ | ☐ | ☐ |
| 5. My job is challenging (sets new goals, is prospective) | ☐ | ☐ | ☐ | ☐ | ☐ |
| 6. I have plenty of freedom how to do my work | ☐ | ☐ | ☐ | ☐ | ☐ |
| 7. I get plenty of opportunities to learn in this job | ☐ | ☐ | ☐ | ☐ | ☐ |
| 8. The facilities/equipment/tools provided are excellent. | ☐ | ☐ | ☐ | ☐ | ☐ |
| 9. I have a lot of support from my boss. | ☐ | ☐ | ☐ | ☐ | ☐ |
| 10. My boss recognizes my work. | ☐ | ☐ | ☐ | ☐ | ☐ |
| 11. 11. The experience i am getting now will be great help in advancing my future career. | ☐ | ☐ | ☐ | ☐ | ☐ |
| 12. I find it easy to keep up with the demands of my job. | ☐ | ☐ | ☐ | ☐ | ☐ |
| 13. I have no problems in achieving balance between my professional and private life. | ☐ | ☐ | ☐ | ☐ | ☐ |
| 14. I like working with my boss. | ☐ | ☐ | ☐ | ☐ | ☐ |
| 15. I get on well with my work colleagues. | ☐ | ☐ | ☐ | ☐ | ☐ |
| 16. I think this organization is a great place to work. | ☐ | ☐ | ☐ | ☐ | ☐ |
| 17. I believe i have a great future in this organization. | ☐ | ☐ | ☐ | ☐ | ☐ |
| 18. I intend to go on working for this organization. | ☐ | ☐ | ☐ | ☐ | ☐ |
| 19. I am happy about the values of this organization – how it conducts its business. | ☐ | ☐ | ☐ | ☐ | ☐ |
| 20. The products/services provided by this organization are excellent. | ☐ | ☐ | ☐ | ☐ | ☐ |

**Acknowledgement**

Funding: This work was supported by the Polish National Agency for Academic Exchange under Grant No. PPI/APM/2019/1/00017/U/00001 project title: *The International Academic Partnership for Generation Z* - 2019-2021.